\documentclass[prl,twocolumn,preprintnumbers,amsmath,amssymb,floatfix]{revtex4-1}
\usepackage{graphicx}
\usepackage{epstopdf}
\usepackage{float}
\usepackage{placeins}
\usepackage{hyperref,graphicx}
\usepackage{dcolumn}
\usepackage{bm}
\usepackage{color}
\usepackage{xcolor}
\usepackage{natbib}
\usepackage{amsmath}
\usepackage{amssymb}
\usepackage{multirow} 
\usepackage{hyperref}
\usepackage{stackengine}
\hypersetup{colorlinks=true, linkcolor=blue, citecolor=red, urlcolor=magenta, pdftitle={2DMOF}, pdfauthor={Saurabh Ghosh}}
\begin{document}
\begin{center}
\begin{large}
\end{large}
\end{center}

\title{Multimode Coupling: A mechanism to engineer multifunctionality at oxide interfaces}
\title{Engineering multifunctionality at oxide interfaces by multimode coupling}
\author{Monirul Shaikh}
\author{Saurabh Ghosh}
\email{saurabhghosh2802@gmail.com}
\affiliation{Department of Physics and Nanotechnology, SRM Institute of Science and Technology, Kattankulathur - 603 203, Tamil Nadu, India}
\begin{abstract}
We employed first-principles density functional theory calculations guided by group-theoretical analysis and demonstrated the control of insulator-metal-insulator transition, polarization and two sublattice magnetization in (LaFeO$_3$)$_1$/(CaFeO$_3$)$_1$ superlattice via. multi structural mode coupling i.e., 'multimode coupling'. We have discovered a polar A-type charge disproportionation mode, Q$_{ACD}$ (analogous to the A-type antiferromagnetic ordering), and found that it couples with the trilinear coupling, $Q_{Tri}$ mode (common in $Pnma$ perovskite oxides and involves three structural modes), and lowers the symmetry further. By tuning the strength of the coupling between the participating modes, the polar metallic phase, polar zero bandgap semiconducting, and polar insulating phases can be obtained. Here, $Q_{Tri}$ switches the polarization direction, whereas, Q$_{ACD}$ can trigger insulator-metal-insulator transition along with the polarization switching. The mechanism is true for any transition metal superlattices constituted with $Pnma$ building blocks and with partially filled $e_g$ or $t_{2g}$ electron(s) at the transition metal sites. 
\end{abstract}
\maketitle
Mode coupling in real materials not only drives structural phase transition but also drives functional properties into the low symmetry phase. The ABO$_3$ perovskite oxides are among the most studied systems owing to their functional properties such as ferroelectricity, weak ferromagnetism, linear magnetoelectricity and many other, primarily driven by structural distortions~\cite{eerenstein2006multiferroic, bousquet2008improper, bostrom2018recipes, pitcher2015tilt, benedek2011}. In $Pnma$ ABO$_3$ oxides, the trilinear coupling, Q$_{Tri}$ $\sim$ Q$_{T}$ $\bigotimes$ Q$_R^+$  $\bigotimes$ Q$_{AFEA}$, between tilt (Q$_{T}$) and in phase rotation (Q$_R^+$) of BO$_6$ octahedra and anti-ferroelectric A-site displacement (Q$_{AFEA}$) mode is responsible for driving the phase transition. By employing layered ordering at the A-site, i.e, by constructing AO and A$^{\prime}$O layer, the anti-ferroelectricity can be tuned into ferroelectricity leading to finite polarization into the system. Here, Q$_{T}$ and Q$_{R^+}$ drive weak ferromagnetism and linear magnetoelectricity, respectively, into the system and the polarization is induced by the trilinear coupling~\cite{ghosh2015, MonirulPhysRevB}. This is known as the hybrid improper ferroelectric (HIF) mechanism~\cite{benedek2011} which is one of the most successful paradigms in designing multiferroic materials where more than one ferroic property can coexist and cross coupled~\cite{rondinelli2012, ghosh2015, MonirulPhysRevB, CmemMat2021, oh2015experimental, pomiro2020first, clarke2021situ}. 
\par
Apart from common distortions  a certain type of structural distortion, termed as a charge disproportionation mode (Q$_{CD}$) has a deterministic role in obtaining insulator-to-metal transition (IMT) \cite{natcommmercy2017}. The Q$_{CD}$ mode is mainly observed in systems those contain charge orderings at the B-sublattice \cite{natcommmercy2017, Jiang2014, rogge2018, Zunger, preziosi2018direct, dominguez2020length}, e.g., perovskite CaFeO$_3$ \cite{Jiang2014, ghosez}. In CaFeO$_3$, electronic phase separation is associated with a charge disproportionation, 2$d^{4+}$ $\rightarrow$ $d^{5+}$ + $d^{3+}$, which is responsible for the band gap openingm\cite{Jiang2014, ghosez}. Similar mechanism follows for other IMT based materials.
In this context, it is proposed that the charge disproportionation and hence the IMT transition can be driven by the intermixing of transition metal ions at the oxide interface and that also can happen in the presence of the polar distortion~\cite{ghosh2017}. 
Further, perturbation, like strain, tunes the local bonding environment leading to the modification of the band structure at the Fermi level~\cite{May}.
\par
In this paper, within (LaFeO$_3$)$_1$/(CaFeO$_3$)$_1$ superlattice (SL) with C-type magnetic configuration, we report a new type of polar charge disproportionation mode Q$_{ACD}$, as here, the charge disproportionation has A-type ordering (in analogy with A-type magnetic orderings). We show a multimode coupling, Q$_M$ $\sim$ Q$_{Tri} \bigotimes$ Q$_{ACD}$ exists between Q$_{Tri}$ and Q$_{ACD}$, which lowers the energy with a lower space group ($Pc$) as compared to the trilinear coupling, Q$_{Tri}$, in perovskite SLs (i,e., $Pmc2_1$)~\cite{PRB2020}.  The electric polarization (\textbf{P}), magnetization (\textbf{M}) and IMT show strong correlation and can be tuned via. strain by modifying the strength of the coupling leading to polar metallic, polar zero bandgap semiconducting, and polar insulating phases. Interestingly, the Q$_{ACD}$ is found to switch the direction of polarization of the system through an insulator-metal-insulator path and orbital ordering plays a pivotal role. Finally, we demonstrated that beyond conventional understanding, the mode coupling at the interface and emergence of associate multifunctionality is applicable for any layered (ABO$_3$)$_1$/(A$^\prime$B$^\prime$O$_3$)$_1$ SLs of transition metal oxides (TMOs) with $Pnma$ symmetry and with unpaired $e_g$ or $t_{2g}$ electrons. 
\par
Density functional theory (DFT) \cite{DFT} calculations are performed for  total energy, geometry optimization and electric polarization using the Berry phase method \cite{Berry} as implemented VASP~\cite{vasp}. The K-integration in the Brillouin zone is performed using $\Gamma$-centered 6 $\times$ 6  $\times$ 4 points using 500.0 eV energy cut-off. We considered the generalized gradient approximation (GGA) augmented by the Hubbard-U corrections (GGA+U) \cite{DFTU, LSDAU} to describe exchange-correlation effects. To consider $d$-$d$ Coulomb interactions, we employed U parameters of 5.0, 3.5, 3.0, 1.0 eV for the Fe-$d$, Cr-$d$, Ru-$d$, Nb-$d$ respectively, and 4.0 eV for Mo-$d$ and Mn-$d$ electrons, while the intra-atomic Hund’s exchange parameter J is kept as 1.0 eV \cite{CmemMat2021}. The exchange-correlation part is estimated by PBEsol functional \cite{PBEsol}. The total energy and Hellman-Feynman force are converged to 0.01 meV and 1 meV/Å, respectively.
\par
\begin{figure}
\centering
\includegraphics[width=\linewidth]{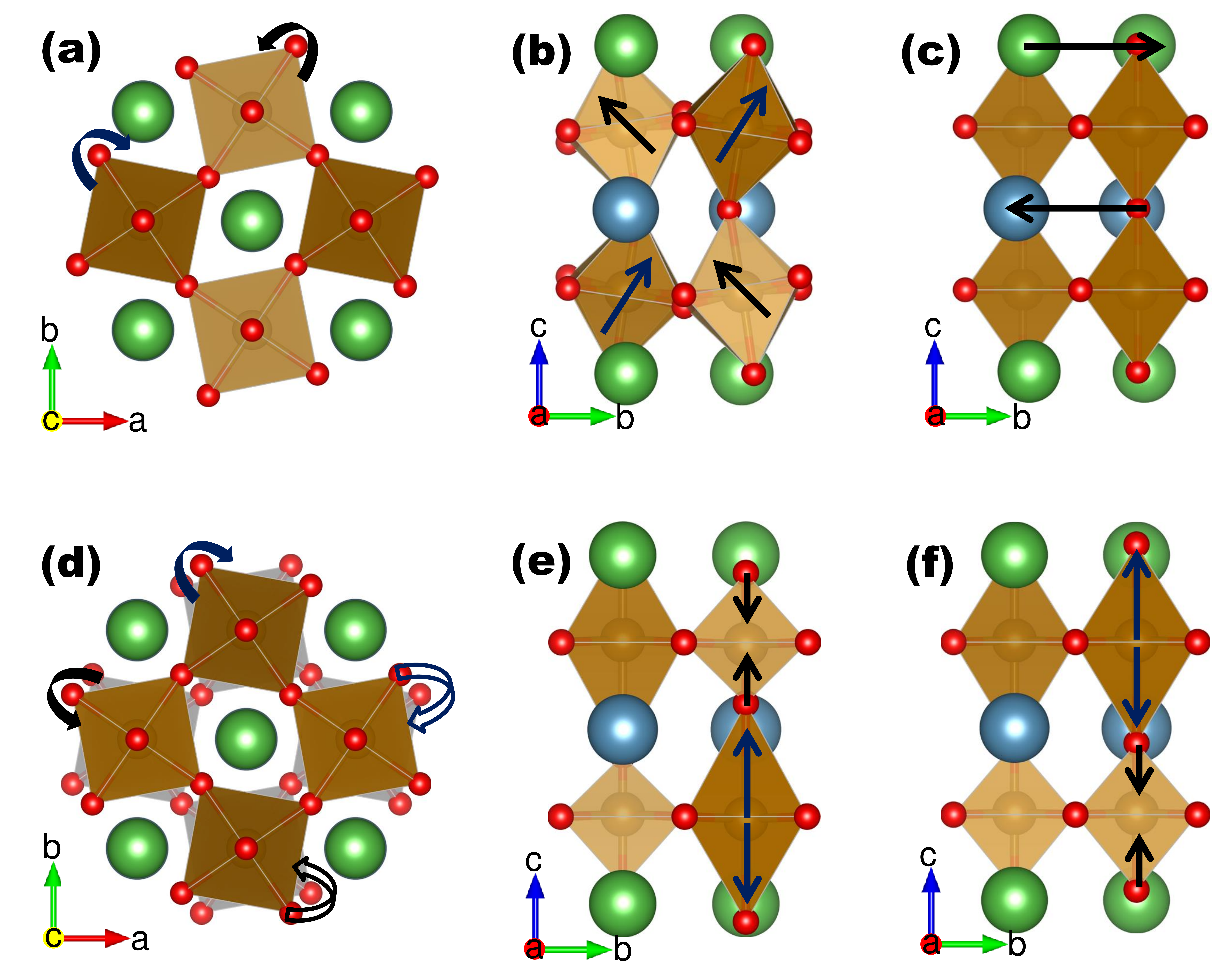}\vspace{-0pt}
\caption {(a) In-phase rotation, a$^0$a$^0$c$^+$ (Q$_R^+$, irrep. M$_2^+$); (b) Tilt a$^-$a$^-$c$^0$ (Q$_T$, irrep. M$_5^-$); (c) Anti-ferroelectric A-site displacement (Q$_{AFEA}$,  irrep. $\Gamma_5^-$), (d) Out-of-phase rotation, a$^0$a$^0$c$^-$ (Q$_R^-$, irrep. M$_4^-$), (e) Charge disproportionation mode analogous to the G-type anti-ferromagnetic ordering (Q$_{GCD}$, irrep. M$_2^-$); and (f) Charge disproportionation mode analogous to the A-type anti-ferromagnetic ordering (Q$_{ACD}$, irrep. $\Gamma_3^-$). The curved arrow with blue (black) is the eye-guide for rotation of FeO$_6$ octahedra in a clock-wise (counter clock-wise) direction. The shape-filled (unfilled) curved arrow highlight the top (bottom) layer rotation of the FeO$_6$ octahedra.}
\label{Modes}
\end{figure}
\begin{figure*}
\centering
\includegraphics[width=\linewidth]{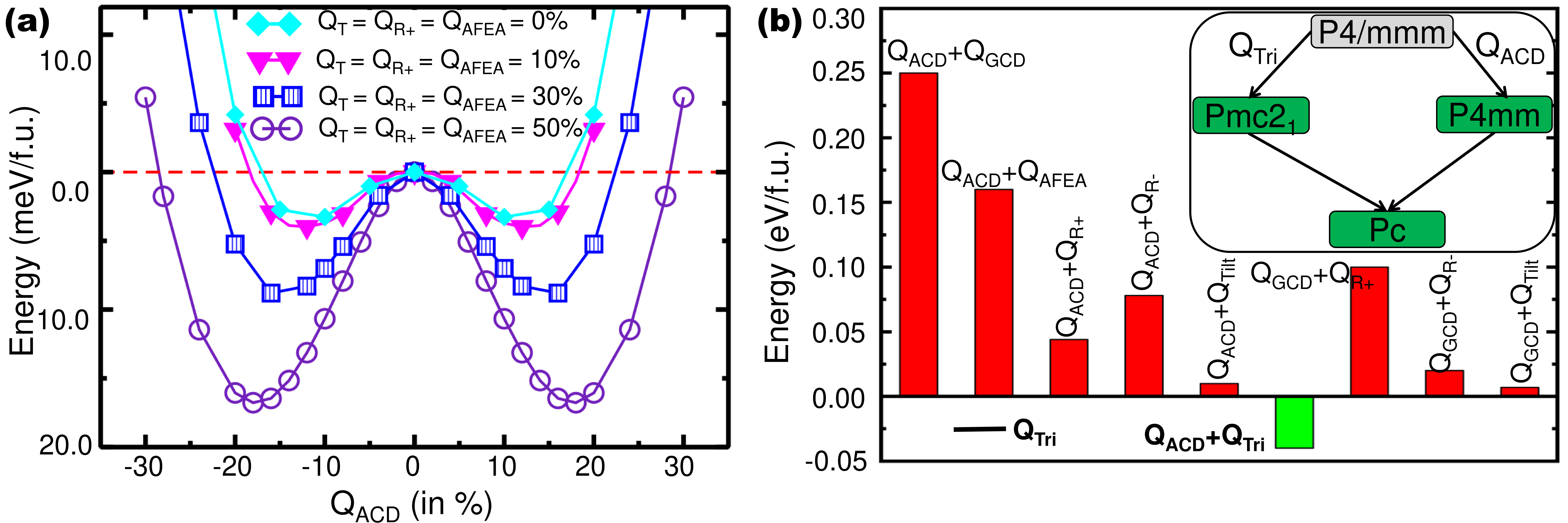}\vspace{-0pt}
\caption {(a) Evolution of energy in terms of Q$_{ACD}$ mode amplitude with fixed amplitudes of Q$_T$, Q$_{R^+}$, and Q$_{AFEA}$ in the C-type antiferromagnetic ordering of $P4/mmm$ phase. (b) Comparative energetic bar diagram for various modes coupling with reference to the trilinear coupling term Q$_{Tri}$; Inset: group-subgroup relationship for a multimode (Q$_M$) coupling, the space group highlighted with green represent non-centrosymmetric space group.} 
\label{Energy}
\end{figure*}

%
We have performed phonon calculations on high symmetry $P4/mmm$ structure of (LaFeO$_3$)$_1$/(CaFeO$_3$)$_1$ SL to get insights into stable and unstable structural modes (see Supplementary Note: I and II for more details). The participating modes that drive the HIF are shown in Figure\ref{Modes} at the top panel. The in-phase rotation Q$_R^+$ (a$^0$a$^0$c$^+$) and tilt Q$_T$ (a$^-$a$^-$c$^0$) are as shown in Figure\ref{Modes}(a) and (b), respectively. The third structural distortion, i.e., anti-ferroelectric A-site displacement mode is shown in Figure\ref{Modes}(c). 
In addition to the HIF modes, we found other three important structural modes. These are out-of-phase rotation Q$_R^-$(a$^0$a$^0$c$^-$) where the in-plane oxygen atoms of FeO$_6$ octahedra in the two successive layers are rotating in the opposite directions as is shown in Figure\ref{Modes}(d). The charge disproportionation mode analogous to the G-type antiferromagnetic configuration (hereafter Q$_{GCD}$) as shown is shown in Figure\ref{Modes}(e). The charge disproportionation mode is analogous to the A-type antiferromagnetic configuration (hereafter Q$_{ACD)}$ leads to a non-centrosymmetric $P4mm$ space group. Unlike Q$_{GCD}$, in Q$_{ACD}$ the FeO$_6$ octahedra in a layer behave alike. This arrangement of charge disproportionated mode reflects in FeO$_6$ octahedra which is in analogy with the A-type antiferromagnetic configuration and is shown in Figure\ref{Modes}(f). The Q$_{ACD}$ mode appeared as imaginary frequency and alone exhibited a double well with small depth. Only coupling with the Q$_{Tri}$ leads to a considerable double well indicating energy gain and strong coupling with Q$_{Tri}$, as shown in 
Figure\ref{Energy}(a). 
\par
\begin{figure}
\centering
\includegraphics[width=\linewidth]{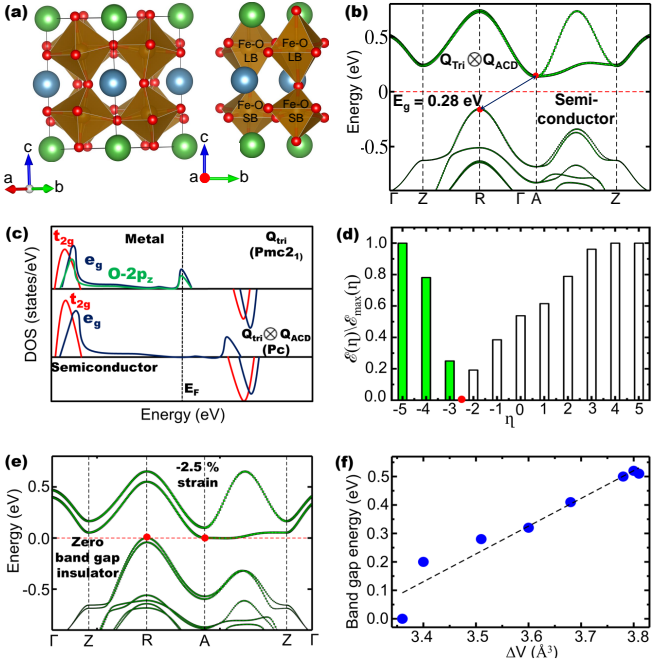}\vspace{-0pt}
\caption {(a) The lowest energy structure $Pc$. Here, LB and SB denote Fe-O long bond and short bond respectively. The green, blue, red, and golden balls represent La-, Ca-, O-, and Fe- atoms, respectively. (b) Band structure of $Pc$, (c) Illustration of insulator-to-metal transition using Crystal Field Splitting, (d) Variation of DOS@E$_F$ or Band Gap as a function of strain $\eta$. Here, $\xi(\eta)$/$\xi_{max}(\eta)$ is defined as the ratio between total DOS@E$_F$ and maximum total DOS@E$_F$ when the system is metallic (filled green), whereas, the same represents the ratio between band gap and maximum band gap when the system is insulating (empty).  The negative sign indicates the compressive strain region, zero (0) denotes without any strain, and right to zero is the tensile strain region. (e) Band structure for zero-band gap insulating phase at compressive 2.5\% strain. (f) The band gap energy is linearly varying with the change in volume, $\Delta$V of the FeO$_6$ octahedra.}
\label{Illustration}
\end{figure}

\par
We then examined how the charge disproportionation modes (i.e., Q$_{GCD}$, Q$_{ACD}$) which are individually responsible for the IMT are coupled to the other dominating distortions. As shown in 
Figure\ref{Energy}(b), among all the modes couplings, we have found that the coupling between Q$_{Tri}$ and Q$_{ACD}$, i.e.; Q$_M$ $\sim$ Q$_{Tri} \bigotimes $Q$_{ACD}$ offers a polar $Pc$ space group and ensures the lowest energy structure (30 meV/f.u. lower than Pmc2$_1$). Here, Q$_M$ represents a multimode coupling term~\cite{ghosh2017}. We have also considered a similar coupling between Q$_{Tri}$ and Q$_{GCD}$ and after geometry relaxation, it turned out that coupling between Q$_{GCD}$ and Q$_{Tri}$ is not stable. Here, to mention as shown in Figure\ref{Energy}(b), both Q$_{Tri}$ and Q$_{ACD}$ can lead to the respective non-centrosymmetric space groups. 
\par
In the presence of Q$_{Tri}$, $Pnma$ LaFeO$_3$ is an antiferromagnetic (G-type) insulator ($d^5$) and $Pnma$ CaFeO$_3$ is a ferromagnetic metal ($d^4$) (Supplementary Note: I). The magnetic ground state of (LaFeO$_3$)$_1$/(CaFeO$_3$)$_1$ superlattice within $Pmc2_1$ symmetry (i.e., in the presence of Q$_{Tri}$) is found to be C-type antiferromagnetic ordering with local magnetic moment on individual Fe is 3.81$\mu_B$ indicating that Fe is in +4 charge state with $d^4$ electronic configuration.  The missing of a hole per formula unit (f.u.) may be attributed to the phenomenon of ligand (O-2$p$) hole recombination~\cite{cho2021}. Upon trilinear mode coupling in (LaFeO$_3$)$_1$/(CaFeO$_3$)$_1$ superlattice, the system is found to be metallic, and due to integration of Q$_{ACD}$ the system emerged as a low bandgap semiconductor with an indirect energy bandgap of 0.28 eV, as shown in Figure~\ref{Illustration} (a-b) within $P_c$ space group. In Figure\ref{Illustration}(a), the LB and SB represent long and short bonds between Fe and O-atoms, respectively, when the Q$_{ACD}$ is coupled with Q$_{Tri}$. The appearance of LB and SB is a clear signature of charge disproportionation. 
\par
Analysis of the partial density of states (PDOS) and orbital projected DOS manifest that Fe-atoms interact antiferromagnetically, while O-2$p$ and Fe-3$d$ interaction is ferromagnetic type. This $p-d$ exchange gives rise to metallic phase due to Fe-$e_{g}$ and O-2$p$  within $Pmc2_1$, as shown in Figure ~\eqref{Illustration}(c) (Supplementary Note: II). Contrary to the ligand-hole recombination for Q$_{Tri}$, in Q$_{M}$ the system goes via. a varying charge states solution. Here, Fe-atoms achieve two different moments, i.e.; 3.47 $\mu_B$ (Fe$^{4+}$ oxidation state, $d^4$) and 4.05 $\mu_B$ (Fe$^{3+}$ oxidation state, $d^5$) satisfying the oxidation of Fe in each block of the superlattice. In this case, the O-2$p_z$ and Fe-3$d_{z^2}$ hybridization is weak, and hence reduced moments are found on O-atoms. In addition, the Fe-e$_{g}$ states pushed below the E$_F$ and the system emerged as a semiconductor within space group \textit{Pc}, as illustrated in Figure\ref{Illustration}(c) (Supplementary Note: II for more details). 
\par
The question is now, how to modulate Q$_{M}$? One possible way to "switch on" the modulation is to apply strain. Upon tensile and compressive strain, (LaFeO$_3$)$_1$/(CaFeO$_3$)$_1$ SL experiences a IMT.  In the compressive region beyond 2.5\% strain, the system is metallic while right from compressive 2.5\% strain to tensile strain region the system behaves as a semiconductor, as shown in Figure ~\ref{Illustration}(d). At compressive 2.5\% strain, it shows a zero-band gap semiconducting phase, as shown in Figure ~\ref{Illustration}(e). The electronic phase separation under strain is found to be a change in volume effect of the FeO$_6$ octahedra of the SL and the reconfiguration of $t_{2g}$ and $e_{g}$ is shown in Figure S5. The energy gap, E$_g$ is found to be varying linearly with the change in volume, $\Delta$V of the FeO$_6$ octahedra analogous to ref.\cite{Jiang2014} and is shown in Figure \ref{Illustration}(f).  
\begin{figure}[t]
\centering
{\includegraphics[width=7.2 cm]{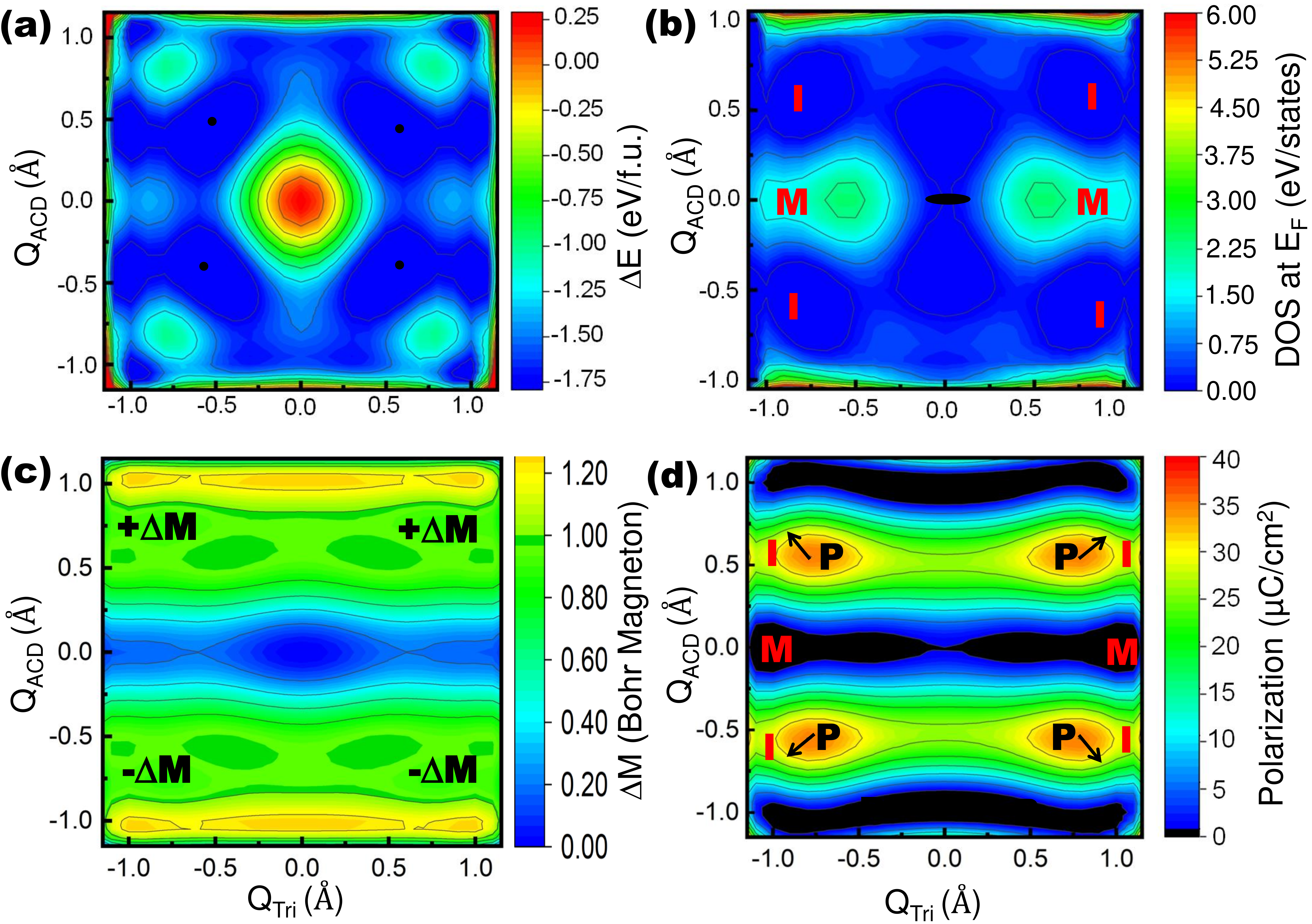}\vspace{-0pt}
} 
\caption {(a) The energy surface contour plot as functions of Q$_{tri}$ and Q$_{ACD}$ by using the Landau theory of phase transition. Subsequently, we have used the same equation for other physical properties to investigate the effect of various modes coupling; (b) density of states at the Fermi level (E$_F$), (c) change in magnetization due to Q$_{ACD}$, and (d) ferroelectric polarization from both Q$_{Tri}$ and Q$_{ACD}$.} 
\label{Contours}
\end{figure}
\par
We fitted the Landau expansion \cite{natcommmercy2017, MonirulPhysRevB, sivadas2021}, for the energy and obtained energy surface contour as a function of Q$_{Tri}$ and Q$_{ACD}$ (Supplementary Note: III for more details). 
\begin {equation}
\begin{split}
f(Q_{Tri}, Q_{ACD}) = a_1Q_{Tri}^2 + a_2Q_{ACD}^2 + a_3Q_{Tri}^2Q_{ACD}^2 \\+ a_4Q_{Tri}^4 + a_5Q_{ACD}^4 + a_6Q_{Tri}^6 + a_7Q_{ACD}^6 \\+ a_8Q_{Tri}^2Q_{ACD}^4+ a_9Q_{Tri}^4Q_{ACD}^2 + a_{10}Q_{Tri}^8 \\+ a_{11}Q_{ACD}^8 + a_{12}Q_{Tri}^4Q_{ACD}^4 + a_{13}Q_{Tri}^6Q_{ACD}^2
\end{split}
\end{equation}
We found four equivalent minima with respect to the high symmetry $P4/mmm$ structure. The most energy gain is found when the amplitudes of both the modes are around $\sim$ 0.5 $\AA$ as shown in Figure\ref{Contours}(a). The coupling between Q$_{Tri}$ and Q$_{ACD}$ lowers the symmetry from metallic $Pmc2_1$ to insulating $Pc$. The question is now how the coupling between Q$_{Tri}$ and Q$_{ACD}$ influences the polarization, magnetization, and density of states (DOS) at the E$_F$. 
\par 
In Figure \ref{Contours}(b), utilizing the same expansion we have shown the variation of density of states at Fermi Level, i.e., DOS$@$E$_F$ as a function of (Q$_{Tri}$, Q$_{ACD}$). The coupling between Q$_{Tri}$ and Q$_{ACD}$ gives rise to insulating and metallic regions. If we consider Q$_{Tri}$ is fixed and switch the Q$_{ACD}$ distortion in the opposite sense, i.e., from $-$Q$_{ACD}$ $\rightarrow$ (Q$_{ACD}$ = 0) $\rightarrow$ $+$ Q$_{ACD}$, we can guide the system from insulating to metallic to again insulating phase. Contrarily, if one goes via. Q$_{Tri}$ mode, it is an insulating phase if Q$_{ACD}$ gives an insulating solution and it is metal if Q$_{ACD}$ is giving a metallic solution. 
The Q$_{ACD}$ mode triggers the charge disproportionation which in turn makes the local magnetic moments on Fe atoms distinguishable by creating two different sublattices.  In Figure\ref{Contours}(c), we have shown the difference in the magnetization i.e., $\Delta$M, between +3 ($d^5$) and +4 ($d^4$) sublattices, as a function of Q$_{Tri}$ and Q$_{ACD}$. The Q$_{ACD}$ mode can reverse the sign of $\Delta$M i.e., the two magnetic sublattices can be interchanged. The $\Delta$M can be as large as 1.2$\mu_B$ per f.u. depending on the strength of Q$_{ACD}$. When optimized to low symmetry $Pc$ phase, the $\Delta$M is found to be 0.58 $\mu_B$ per f.u. Considering Figure \ref{Contours}(b) and (c), a clear correlation between DOS$@$E$_F$ and $\Delta$M can be observed.
\par
The Q$_{AFEA}$ mode give rise to the finite polarization \textbf {P} when La/Ca layered ordering has been established within Q$_M$ $\sim$ Q$_{Tri}$ $\bigotimes$ Q$_{ACD}$ and Q$_{ACD}$ ensures an insulating phase. While varying the Q$_{Tri}$, we have first fixed the Q$_{AFEA}$ at various magnitudes and by selective dynamics, Q$_{T}$ and Q$_R^+$ distortions are allowed to relax. This establishes the fact that Q$_{AFEA}$ (or Q$_P$) $\sim$ Q$_{Tri}$. In Figure ~\ref{Contours}(d) we have shown the variation of \textbf {P} as a $f=$ (Q$_{Tri}$, Q$_{ACD}$). When Q$_{ACD}$ = 0, the polarization can not be measured as the system remains in the metallic region. But when both Q$_{Tri}$ and Q$_{ACD}$ mods are around $\sim$ 0.5 $\AA$, it shows a polarization as large as 40 $\mu$C/cm$^2$ in the insulating phase. In the case of the optimized $Pc$ phase, the total polarization is computed as  36.38 $\mu$C/cm$^2$. Interestingly, we noticed that the directions of the polarization at four equivalent minima are different. This is due to the fact that both Q$_{AFEA}$ and Q$_{ACD}$ are polar modes which give rise to the finite polarization along Cartesian $y$, P$_y$ (crystallographic $b$) and $z$, P$_z$ (Crystallographic $c$) directions, respectively. Along the Q$_{AFE}$ (i.e, Q$_{Tri}$) the P$_y$ can be reversed whereas along Q$_{ACD}$, P$_z$ component can be reversed. The four polarization domains of (P$_y$, P$_z$) as $++$, $-+$, $+-$ and $--$ are possible considering coupling between Q$_{Tri}$ and Q$_{ACD}$.  Along the Q$_{Tri}$, only polarization switching can be observed. Interestingly, through Q$_{ACD}$ system undergoes insulator-metal-insulator transition along with the polarization switching.  
\par
\begin{table}[b]
\caption{\label{tab:1} Bandgap (B.G), polarization (P), and change in magnetization ($\Delta$M) for 3$d$-4$d$ (LaBO$_3$)$_1$/(CaB$^\prime$O$_3$)$_1$ superlattices (SL). Here, M indicates metallic phase.}
\begin{ruledtabular}
\begin{tabular}{ccccccc}
\textrm{(LaBO$_3$)$_1$/(CaB$^\prime$O$_3$)$_1$} & \textrm{B.G} & \textrm{P} & \textrm{$\Delta$M} \\
\textrm{B(3$d$)/B$^\prime$(4$d$)} & \textrm{(eV)} & \textrm{($\mu$C/cm$^2$)} & \textrm{($\mu_B$)} \\
\colrule
Cr/Nb & $M$ & $M$ & 1.71\\
Cr/Mo& 1.52 & 23.74 & 1.01\\
Cr/Ru& 1.40 & 26.60 & 1.16\\
\colrule
Mn/Nb & $M$ & $M$ & 2.62\\
Mn/Mo& 0.98 & 22.62 & 1.81\\
Mn/Ru & 0.03 & 27.39 & 2.01\\
\colrule
Fe/Nb & $M$ & $M$ & 3.30\\
Fe/Mo  & 1.20 & 21.15 & 2.25\\
Fe/Ru & 0.78 & 24.40 & 2.49\\
\colrule
(LaFeO$_3$)$_1$/(CaFeO$_3$)$_1$ & 0.28 & 36.38 & 0.58\\
\end{tabular}
\end{ruledtabular}
\label{propsSL}
\end{table}
\par
In (LaFeO$_3$)$_1$/(CaFeO$_3$)$_1$ SL, the emergent properties i.e, IMT, finite \textbf{P} and \textbf{M} and cross coupling between them depends on the strength of the coupling coefficients (i.e, $a_i$'s where $i$= 1,2,3 ...) as given in Equation 1. These coupling coefficients are materials specific but the effect is universal. To prove the concept of universality of multimode coupling, we have constructed artificial heterostrcuture superlattices of (LaBO$_3$)$_1$/(CaB$^\prime$O$_3$)$_1$, where B in (+3 oxidation state)= Cr(3$d^3$), Mn (3$d^4$, $e^1_g$), Fe(3$d^5$) and B$^\prime$ (in +4 oxidation state) = Nb(4$d^1$,$t^1_{2g}$), Mo(4$d^2$, $t^2_{2g}$) and Ru(4$d^4$, $e^1_g$). We have started with the high symmetry  $P4/mmm$ phase and invoke the trilinear coupling term Q$_{Tri}$ into the system. A complete relaxation without symmetry constrain shows that the Q$_{ACD}$ mode clubbing into the trilinear coupling term. This shows universal coupling between Q$_{Tri}$ and Q$_{ACD}$.  
The properties of (LaBO$_3$)$_1$/(CaB$^\prime$O$_3$)$_1$  SLs are tabulated in Table~\ref{propsSL}. All  the Nb-based compounds are metallic whereas all the Mo-and Ru-based compounds are found to be insulating with with band gap ranging from 0.03 eV - 1.52 eV, with maximum for (LaCrO$_3$)$_1$/(CaMoO$_3$)$_1$. Thus, we have identified both polar-metal and polar insulators with large ferroelectric (or, 'ferroelectric like' in the case of metals \cite{ghosh2017}) distortion. The change in $\Delta M$ follows a linear trend for a fixed 4$d$ and for (LaFeO$_3$)$_1$/(CaNbO$_3$)$_1$ the value of $\Delta M$ is maximum as 3.5 $\mu_B$. Total polarization values are found to be within the range of 21.5 - 27.39 $\mu$C/cm$^2$ and largest for (LaMnO$_3$)$_1$/(CaRuO$_3$)$_1$ heterostructure (polar vector Figure S7). 
\par
To conclude, we have proposed a mechanism of multimode coupling through which functional properties such as metal-to-insulator transition, polarisation and magnetization can be engineered at the (LaFeO$_3$)$_1$/(CaFeO$_3$)$_1$ oxide interface. The polarization is driven into the system by trilinear coupling $Q_{Tri}$, via. Q$_{AFEA}$. The Q$_{ACD}$ mode drives metal-to-insulator transition and polarization. By tuning the coupling between $Q_{Tri}$ and Q$_{ACD}$, polar metallic phase, polar zero band gap semiconducting, and polar insulating phases can be obtained. The in-plane and out-of-plane polarization is induced into the system via. $Q_{AFEA}$ and Q$_{ACD}$ distortions, respectively. Through Q$_{ACD}$ path, the system undergoes insulator-metal-insulator transition along the polarization switching. The coupling mechanism between $Q_{Tri}$ and Q$_{ACD}$ and emergence of associated properties can be obtained for other superlattices with  $Pnma$ building block and partially filled $e_g$ or $t_{2g}$ electron(s) at the B or B$^{\prime}$ sites. 
\par
M.S. acknowledges INSPIRE fellowship of Department of Science and Technology,  New Delhi-110 016, Government of India. S.G. acknowledges DST-SERB Core Research Grant, File No. CRG/2018/001728 (2019-2022) for funding. M.S. and S.G. are grateful to SRM Institute of Science and Technology, Chennai, India for computational resources.

\begin{thebibliography}{30}%
\makeatletter
\providecommand \@ifxundefined [1]{%
 \@ifx{#1\undefined}
}%
\providecommand \@ifnum [1]{%
 \ifnum #1\expandafter \@firstoftwo
 \else \expandafter \@secondoftwo
 \fi
}%
\providecommand \@ifx [1]{%
 \ifx #1\expandafter \@firstoftwo
 \else \expandafter \@secondoftwo
 \fi
}%
\providecommand \natexlab [1]{#1}%
\providecommand \enquote  [1]{``#1''}%
\providecommand \bibnamefont  [1]{#1}%
\providecommand \bibfnamefont [1]{#1}%
\providecommand \citenamefont [1]{#1}%
\providecommand \href@noop [0]{\@secondoftwo}%
\providecommand \href [0]{\begingroup \@sanitize@url \@href}%
\providecommand \@href[1]{\@@startlink{#1}\@@href}%
\providecommand \@@href[1]{\endgroup#1\@@endlink}%
\providecommand \@sanitize@url [0]{\catcode `\\12\catcode `\$12\catcode
  `\&12\catcode `\#12\catcode `\^12\catcode `\_12\catcode `\%12\relax}%
\providecommand \@@startlink[1]{}%
\providecommand \@@endlink[0]{}%
\providecommand \url  [0]{\begingroup\@sanitize@url \@url }%
\providecommand \@url [1]{\endgroup\@href {#1}{\urlprefix }}%
\providecommand \urlprefix  [0]{URL }%
\providecommand \Eprint [0]{\href }%
\providecommand \doibase [0]{http://dx.doi.org/}%
\providecommand \selectlanguage [0]{\@gobble}%
\providecommand \bibinfo  [0]{\@secondoftwo}%
\providecommand \bibfield  [0]{\@secondoftwo}%
\providecommand \translation [1]{[#1]}%
\providecommand \BibitemOpen [0]{}%
\providecommand \bibitemStop [0]{}%
\providecommand \bibitemNoStop [0]{.\EOS\space}%
\providecommand \EOS [0]{\spacefactor3000\relax}%
\providecommand \BibitemShut  [1]{\csname bibitem#1\endcsname}%
\let\auto@bib@innerbib\@empty
\bibitem [{\citenamefont {Eerenstein}\ \emph {et~al.}(2006)\citenamefont
  {Eerenstein}, \citenamefont {Mathur},\ and\ \citenamefont
  {Scott}}]{eerenstein2006multiferroic}%
  \BibitemOpen
  \bibfield  {author} {\bibinfo {author} {\bibfnamefont {W.}~\bibnamefont
  {Eerenstein}}, \bibinfo {author} {\bibfnamefont {N.}~\bibnamefont {Mathur}},
  \ and\ \bibinfo {author} {\bibfnamefont {J.~F.}\ \bibnamefont {Scott}},\
  }\href {\doibase 10.1038/nature05023} {\bibfield  {journal} {\bibinfo
  {journal} {nature}\ }\textbf {\bibinfo {volume} {442}},\ \bibinfo {pages}
  {759} (\bibinfo {year} {2006})}\BibitemShut {NoStop}%
\bibitem [{\citenamefont {Bousquet}\ \emph {et~al.}(2008)\citenamefont
  {Bousquet}, \citenamefont {Dawber}, \citenamefont {Stucki}, \citenamefont
  {Lichtensteiger}, \citenamefont {Hermet}, \citenamefont {Gariglio},
  \citenamefont {Triscone},\ and\ \citenamefont
  {Ghosez}}]{bousquet2008improper}%
  \BibitemOpen
  \bibfield  {author} {\bibinfo {author} {\bibfnamefont {E.}~\bibnamefont
  {Bousquet}}, \bibinfo {author} {\bibfnamefont {M.}~\bibnamefont {Dawber}},
  \bibinfo {author} {\bibfnamefont {N.}~\bibnamefont {Stucki}}, \bibinfo
  {author} {\bibfnamefont {C.}~\bibnamefont {Lichtensteiger}}, \bibinfo
  {author} {\bibfnamefont {P.}~\bibnamefont {Hermet}}, \bibinfo {author}
  {\bibfnamefont {S.}~\bibnamefont {Gariglio}}, \bibinfo {author}
  {\bibfnamefont {J.-M.}\ \bibnamefont {Triscone}}, \ and\ \bibinfo {author}
  {\bibfnamefont {P.}~\bibnamefont {Ghosez}},\ }\href {\doibase
  10.1038/nature06817} {\bibfield  {journal} {\bibinfo  {journal} {Nature}\
  }\textbf {\bibinfo {volume} {452}},\ \bibinfo {pages} {732} (\bibinfo {year}
  {2008})}\BibitemShut {NoStop}%
\bibitem [{\citenamefont {Bostr{\"o}m}\ \emph {et~al.}(2018)\citenamefont
  {Bostr{\"o}m}, \citenamefont {Senn},\ and\ \citenamefont
  {Goodwin}}]{bostrom2018recipes}%
  \BibitemOpen
  \bibfield  {author} {\bibinfo {author} {\bibfnamefont {H.~L.}\ \bibnamefont
  {Bostr{\"o}m}}, \bibinfo {author} {\bibfnamefont {M.~S.}\ \bibnamefont
  {Senn}}, \ and\ \bibinfo {author} {\bibfnamefont {A.~L.}\ \bibnamefont
  {Goodwin}},\ }\href {\doibase 10.1038/s41467-018-04764-x} {\bibfield
  {journal} {\bibinfo  {journal} {Nature communications}\ }\textbf {\bibinfo
  {volume} {9}},\ \bibinfo {pages} {1} (\bibinfo {year} {2018})}\BibitemShut
  {NoStop}%
\bibitem [{\citenamefont {Pitcher}\ \emph {et~al.}(2015)\citenamefont
  {Pitcher}, \citenamefont {Mandal}, \citenamefont {Dyer}, \citenamefont
  {Alaria}, \citenamefont {Borisov}, \citenamefont {Niu}, \citenamefont
  {Claridge},\ and\ \citenamefont {Rosseinsky}}]{pitcher2015tilt}%
  \BibitemOpen
  \bibfield  {author} {\bibinfo {author} {\bibfnamefont {M.~J.}\ \bibnamefont
  {Pitcher}}, \bibinfo {author} {\bibfnamefont {P.}~\bibnamefont {Mandal}},
  \bibinfo {author} {\bibfnamefont {M.~S.}\ \bibnamefont {Dyer}}, \bibinfo
  {author} {\bibfnamefont {J.}~\bibnamefont {Alaria}}, \bibinfo {author}
  {\bibfnamefont {P.}~\bibnamefont {Borisov}}, \bibinfo {author} {\bibfnamefont
  {H.}~\bibnamefont {Niu}}, \bibinfo {author} {\bibfnamefont {J.~B.}\
  \bibnamefont {Claridge}}, \ and\ \bibinfo {author} {\bibfnamefont {M.~J.}\
  \bibnamefont {Rosseinsky}},\ }\href {\doibase 10.1126/science.1262118}
  {\bibfield  {journal} {\bibinfo  {journal} {Science}\ }\textbf {\bibinfo
  {volume} {347}},\ \bibinfo {pages} {420} (\bibinfo {year}
  {2015})}\BibitemShut {NoStop}%
\bibitem [{\citenamefont {Benedek}\ and\ \citenamefont
  {Fennie}(2011)}]{benedek2011}%
  \BibitemOpen
  \bibfield  {author} {\bibinfo {author} {\bibfnamefont {N.~A.}\ \bibnamefont
  {Benedek}}\ and\ \bibinfo {author} {\bibfnamefont {C.~J.}\ \bibnamefont
  {Fennie}},\ }\href {\doibase 10.1103/PhysRevLett.106.107204} {\bibfield
  {journal} {\bibinfo  {journal} {Phys. Rev. Lett.}\ }\textbf {\bibinfo
  {volume} {106}},\ \bibinfo {pages} {107204} (\bibinfo {year}
  {2011})}\BibitemShut {NoStop}%
\bibitem [{\citenamefont {Ghosh}\ \emph {et~al.}(2015)\citenamefont {Ghosh},
  \citenamefont {Das},\ and\ \citenamefont {Fennie}}]{ghosh2015}%
  \BibitemOpen
  \bibfield  {author} {\bibinfo {author} {\bibfnamefont {S.}~\bibnamefont
  {Ghosh}}, \bibinfo {author} {\bibfnamefont {H.}~\bibnamefont {Das}}, \ and\
  \bibinfo {author} {\bibfnamefont {C.~J.}\ \bibnamefont {Fennie}},\ }\href
  {\doibase 10.1103/PhysRevB.92.184112} {\bibfield  {journal} {\bibinfo
  {journal} {Phys. Rev. B}\ }\textbf {\bibinfo {volume} {92}},\ \bibinfo
  {pages} {184112} (\bibinfo {year} {2015})}\BibitemShut {NoStop}%
\bibitem [{\citenamefont {Shaikh}\ \emph
  {et~al.}(2020{\natexlab{a}})\citenamefont {Shaikh}, \citenamefont
  {Karmakar},\ and\ \citenamefont {Ghosh}}]{MonirulPhysRevB}%
  \BibitemOpen
  \bibfield  {author} {\bibinfo {author} {\bibfnamefont {M.}~\bibnamefont
  {Shaikh}}, \bibinfo {author} {\bibfnamefont {M.}~\bibnamefont {Karmakar}}, \
  and\ \bibinfo {author} {\bibfnamefont {S.}~\bibnamefont {Ghosh}},\ }\href
  {\doibase 10.1103/PhysRevB.101.054101} {\bibfield  {journal} {\bibinfo
  {journal} {Phys. Rev. B}\ }\textbf {\bibinfo {volume} {101}},\ \bibinfo
  {pages} {054101} (\bibinfo {year} {2020}{\natexlab{a}})}\BibitemShut
  {NoStop}%
\bibitem [{\citenamefont {Rondinelli}\ and\ \citenamefont
  {Fennie}(2012)}]{rondinelli2012}%
  \BibitemOpen
  \bibfield  {author} {\bibinfo {author} {\bibfnamefont {J.~M.}\ \bibnamefont
  {Rondinelli}}\ and\ \bibinfo {author} {\bibfnamefont {C.~J.}\ \bibnamefont
  {Fennie}},\ }\href {\doibase 10.1002/adma.201104674} {\bibfield  {journal}
  {\bibinfo  {journal} {Advanced Materials}\ }\textbf {\bibinfo {volume}
  {24}},\ \bibinfo {pages} {1918} (\bibinfo {year} {2012})}\BibitemShut
  {NoStop}%
\bibitem [{\citenamefont {Shaikh}\ \emph {et~al.}(2021)\citenamefont {Shaikh},
  \citenamefont {Fathima}, \citenamefont {Swamynadhan}, \citenamefont {Das},\
  and\ \citenamefont {Ghosh}}]{CmemMat2021}%
  \BibitemOpen
  \bibfield  {author} {\bibinfo {author} {\bibfnamefont {M.}~\bibnamefont
  {Shaikh}}, \bibinfo {author} {\bibfnamefont {A.}~\bibnamefont {Fathima}},
  \bibinfo {author} {\bibfnamefont {M.}~\bibnamefont {Swamynadhan}}, \bibinfo
  {author} {\bibfnamefont {H.}~\bibnamefont {Das}}, \ and\ \bibinfo {author}
  {\bibfnamefont {S.}~\bibnamefont {Ghosh}},\ }\href {\doibase
  10.1021/acs.chemmater.0c02976} {\bibfield  {journal} {\bibinfo  {journal}
  {Chemistry of Materials}\ }\textbf {\bibinfo {volume} {33}},\ \bibinfo
  {pages} {1594} (\bibinfo {year} {2021})}\BibitemShut {NoStop}%
\bibitem [{\citenamefont {Oh}\ \emph {et~al.}(2015)\citenamefont {Oh},
  \citenamefont {Luo}, \citenamefont {Huang}, \citenamefont {Wang},\ and\
  \citenamefont {Cheong}}]{oh2015experimental}%
  \BibitemOpen
  \bibfield  {author} {\bibinfo {author} {\bibfnamefont {Y.~S.}\ \bibnamefont
  {Oh}}, \bibinfo {author} {\bibfnamefont {X.}~\bibnamefont {Luo}}, \bibinfo
  {author} {\bibfnamefont {F.-T.}\ \bibnamefont {Huang}}, \bibinfo {author}
  {\bibfnamefont {Y.}~\bibnamefont {Wang}}, \ and\ \bibinfo {author}
  {\bibfnamefont {S.-W.}\ \bibnamefont {Cheong}},\ }\href {\doibase
  10.1038/nmat4168} {\bibfield  {journal} {\bibinfo  {journal} {Nature
  materials}\ }\textbf {\bibinfo {volume} {14}},\ \bibinfo {pages} {407}
  (\bibinfo {year} {2015})}\BibitemShut {NoStop}%
\bibitem [{\citenamefont {Pomiro}\ \emph {et~al.}(2020)\citenamefont {Pomiro},
  \citenamefont {Ablitt}, \citenamefont {Bristowe}, \citenamefont {Mostofi},
  \citenamefont {Won}, \citenamefont {Cheong},\ and\ \citenamefont
  {Senn}}]{pomiro2020first}%
  \BibitemOpen
  \bibfield  {author} {\bibinfo {author} {\bibfnamefont {F.}~\bibnamefont
  {Pomiro}}, \bibinfo {author} {\bibfnamefont {C.}~\bibnamefont {Ablitt}},
  \bibinfo {author} {\bibfnamefont {N.~C.}\ \bibnamefont {Bristowe}}, \bibinfo
  {author} {\bibfnamefont {A.~A.}\ \bibnamefont {Mostofi}}, \bibinfo {author}
  {\bibfnamefont {C.}~\bibnamefont {Won}}, \bibinfo {author} {\bibfnamefont
  {S.-W.}\ \bibnamefont {Cheong}}, \ and\ \bibinfo {author} {\bibfnamefont
  {M.~S.}\ \bibnamefont {Senn}},\ }\href {\doibase 10.1103/PhysRevB.102.014101}
  {\bibfield  {journal} {\bibinfo  {journal} {Phys. Rev. B}\ }\textbf {\bibinfo
  {volume} {102}},\ \bibinfo {pages} {014101} (\bibinfo {year}
  {2020})}\BibitemShut {NoStop}%
\bibitem [{\citenamefont {Clarke}\ \emph {et~al.}(2021)\citenamefont {Clarke},
  \citenamefont {Ablitt}, \citenamefont {Daniels}, \citenamefont {Checchia},\
  and\ \citenamefont {Senn}}]{clarke2021situ}%
  \BibitemOpen
  \bibfield  {author} {\bibinfo {author} {\bibfnamefont {G.}~\bibnamefont
  {Clarke}}, \bibinfo {author} {\bibfnamefont {C.}~\bibnamefont {Ablitt}},
  \bibinfo {author} {\bibfnamefont {J.}~\bibnamefont {Daniels}}, \bibinfo
  {author} {\bibfnamefont {S.}~\bibnamefont {Checchia}}, \ and\ \bibinfo
  {author} {\bibfnamefont {M.~S.}\ \bibnamefont {Senn}},\ }\href {\doibase
  10.1107/S1600576721001096} {\bibfield  {journal} {\bibinfo  {journal}
  {Journal of applied crystallography}\ }\textbf {\bibinfo {volume} {54}}
  (\bibinfo {year} {2021}),\ 10.1107/S1600576721001096}\BibitemShut {NoStop}%
\bibitem [{\citenamefont {Mercy}\ \emph {et~al.}(2017)\citenamefont {Mercy},
  \citenamefont {Bieder}, \citenamefont {{\'I}{\~n}iguez},\ and\ \citenamefont
  {Ghosez}}]{natcommmercy2017}%
  \BibitemOpen
  \bibfield  {author} {\bibinfo {author} {\bibfnamefont {A.}~\bibnamefont
  {Mercy}}, \bibinfo {author} {\bibfnamefont {J.}~\bibnamefont {Bieder}},
  \bibinfo {author} {\bibfnamefont {J.}~\bibnamefont {{\'I}{\~n}iguez}}, \ and\
  \bibinfo {author} {\bibfnamefont {P.}~\bibnamefont {Ghosez}},\ }\href
  {\doibase 10.1038/s41467-017-01811-x} {\bibfield  {journal} {\bibinfo
  {journal} {Nature communications}\ }\textbf {\bibinfo {volume} {8}},\
  \bibinfo {pages} {1} (\bibinfo {year} {2017})}\BibitemShut {NoStop}%
\bibitem [{\citenamefont {Jiang}\ \emph {et~al.}(2014)\citenamefont {Jiang},
  \citenamefont {Saldana-Greco}, \citenamefont {Schick},\ and\ \citenamefont
  {Rappe}}]{Jiang2014}%
  \BibitemOpen
  \bibfield  {author} {\bibinfo {author} {\bibfnamefont {L.}~\bibnamefont
  {Jiang}}, \bibinfo {author} {\bibfnamefont {D.}~\bibnamefont
  {Saldana-Greco}}, \bibinfo {author} {\bibfnamefont {J.~T.}\ \bibnamefont
  {Schick}}, \ and\ \bibinfo {author} {\bibfnamefont {A.~M.}\ \bibnamefont
  {Rappe}},\ }\href {\doibase 10.1103/PhysRevB.89.235106} {\bibfield  {journal}
  {\bibinfo  {journal} {Phys. Rev. B}\ }\textbf {\bibinfo {volume} {89}},\
  \bibinfo {pages} {235106} (\bibinfo {year} {2014})}\BibitemShut {NoStop}%
\bibitem [{\citenamefont {Rogge}\ \emph
  {et~al.}(2018{\natexlab{a}})\citenamefont {Rogge}, \citenamefont
  {Chandrasena}, \citenamefont {Cammarata}, \citenamefont {Green},
  \citenamefont {Shafer}, \citenamefont {Lefler}, \citenamefont {Huon},
  \citenamefont {Arab}, \citenamefont {Arenholz}, \citenamefont {Lee},
  \citenamefont {Lee}, \citenamefont {Nem\ifmmode~\check{s}\else
  \v{s}\fi{}\'ak}, \citenamefont {Rondinelli}, \citenamefont {Gray},\ and\
  \citenamefont {May}}]{rogge2018}%
  \BibitemOpen
  \bibfield  {author} {\bibinfo {author} {\bibfnamefont {P.~C.}\ \bibnamefont
  {Rogge}}, \bibinfo {author} {\bibfnamefont {R.~U.}\ \bibnamefont
  {Chandrasena}}, \bibinfo {author} {\bibfnamefont {A.}~\bibnamefont
  {Cammarata}}, \bibinfo {author} {\bibfnamefont {R.~J.}\ \bibnamefont
  {Green}}, \bibinfo {author} {\bibfnamefont {P.}~\bibnamefont {Shafer}},
  \bibinfo {author} {\bibfnamefont {B.~M.}\ \bibnamefont {Lefler}}, \bibinfo
  {author} {\bibfnamefont {A.}~\bibnamefont {Huon}}, \bibinfo {author}
  {\bibfnamefont {A.}~\bibnamefont {Arab}}, \bibinfo {author} {\bibfnamefont
  {E.}~\bibnamefont {Arenholz}}, \bibinfo {author} {\bibfnamefont {H.~N.}\
  \bibnamefont {Lee}}, \bibinfo {author} {\bibfnamefont {T.-L.}\ \bibnamefont
  {Lee}}, \bibinfo {author} {\bibfnamefont {S.}~\bibnamefont
  {Nem\ifmmode~\check{s}\else \v{s}\fi{}\'ak}}, \bibinfo {author}
  {\bibfnamefont {J.~M.}\ \bibnamefont {Rondinelli}}, \bibinfo {author}
  {\bibfnamefont {A.~X.}\ \bibnamefont {Gray}}, \ and\ \bibinfo {author}
  {\bibfnamefont {S.~J.}\ \bibnamefont {May}},\ }\href {\doibase
  10.1103/PhysRevMaterials.2.015002} {\bibfield  {journal} {\bibinfo  {journal}
  {Phys. Rev. Materials}\ }\textbf {\bibinfo {volume} {2}},\ \bibinfo {pages}
  {015002} (\bibinfo {year} {2018}{\natexlab{a}})}\BibitemShut {NoStop}%
\bibitem [{\citenamefont {Dalpian}\ \emph {et~al.}(2018)\citenamefont
  {Dalpian}, \citenamefont {Liu}, \citenamefont {Varignon}, \citenamefont
  {Bibes},\ and\ \citenamefont {Zunger}}]{Zunger}%
  \BibitemOpen
  \bibfield  {author} {\bibinfo {author} {\bibfnamefont {G.~M.}\ \bibnamefont
  {Dalpian}}, \bibinfo {author} {\bibfnamefont {Q.}~\bibnamefont {Liu}},
  \bibinfo {author} {\bibfnamefont {J.}~\bibnamefont {Varignon}}, \bibinfo
  {author} {\bibfnamefont {M.}~\bibnamefont {Bibes}}, \ and\ \bibinfo {author}
  {\bibfnamefont {A.}~\bibnamefont {Zunger}},\ }\href {\doibase
  10.1103/PhysRevB.98.075135} {\bibfield  {journal} {\bibinfo  {journal} {Phys.
  Rev. B}\ }\textbf {\bibinfo {volume} {98}},\ \bibinfo {pages} {075135}
  (\bibinfo {year} {2018})}\BibitemShut {NoStop}%
\bibitem [{\citenamefont {Preziosi}\ \emph {et~al.}(2018)\citenamefont
  {Preziosi}, \citenamefont {Lopez-Mir}, \citenamefont {Li}, \citenamefont
  {Cornelissen}, \citenamefont {Lee}, \citenamefont {Trier}, \citenamefont
  {Bouzehouane}, \citenamefont {Valencia}, \citenamefont {Gloter},
  \citenamefont {Barth{\'e}l{\'e}my} \emph {et~al.}}]{preziosi2018direct}%
  \BibitemOpen
  \bibfield  {author} {\bibinfo {author} {\bibfnamefont {D.}~\bibnamefont
  {Preziosi}}, \bibinfo {author} {\bibfnamefont {L.}~\bibnamefont {Lopez-Mir}},
  \bibinfo {author} {\bibfnamefont {X.}~\bibnamefont {Li}}, \bibinfo {author}
  {\bibfnamefont {T.}~\bibnamefont {Cornelissen}}, \bibinfo {author}
  {\bibfnamefont {J.~H.}\ \bibnamefont {Lee}}, \bibinfo {author} {\bibfnamefont
  {F.}~\bibnamefont {Trier}}, \bibinfo {author} {\bibfnamefont
  {K.}~\bibnamefont {Bouzehouane}}, \bibinfo {author} {\bibfnamefont
  {S.}~\bibnamefont {Valencia}}, \bibinfo {author} {\bibfnamefont
  {A.}~\bibnamefont {Gloter}}, \bibinfo {author} {\bibfnamefont
  {A.}~\bibnamefont {Barth{\'e}l{\'e}my}},  \emph {et~al.},\ }\href {\doibase
  10.1021/acs.nanolett.7b04728} {\bibfield  {journal} {\bibinfo  {journal}
  {Nano letters}\ }\textbf {\bibinfo {volume} {18}},\ \bibinfo {pages} {2226}
  (\bibinfo {year} {2018})}\BibitemShut {NoStop}%
\bibitem [{\citenamefont {Dom{\'\i}nguez}\ \emph {et~al.}(2020)\citenamefont
  {Dom{\'\i}nguez}, \citenamefont {Georgescu}, \citenamefont {Mundet},
  \citenamefont {Zhang}, \citenamefont {Fowlie}, \citenamefont {Mercy},
  \citenamefont {Waelchli}, \citenamefont {Catalano}, \citenamefont
  {Alexander}, \citenamefont {Ghosez} \emph {et~al.}}]{dominguez2020length}%
  \BibitemOpen
  \bibfield  {author} {\bibinfo {author} {\bibfnamefont {C.}~\bibnamefont
  {Dom{\'\i}nguez}}, \bibinfo {author} {\bibfnamefont {A.~B.}\ \bibnamefont
  {Georgescu}}, \bibinfo {author} {\bibfnamefont {B.}~\bibnamefont {Mundet}},
  \bibinfo {author} {\bibfnamefont {Y.}~\bibnamefont {Zhang}}, \bibinfo
  {author} {\bibfnamefont {J.}~\bibnamefont {Fowlie}}, \bibinfo {author}
  {\bibfnamefont {A.}~\bibnamefont {Mercy}}, \bibinfo {author} {\bibfnamefont
  {A.}~\bibnamefont {Waelchli}}, \bibinfo {author} {\bibfnamefont
  {S.}~\bibnamefont {Catalano}}, \bibinfo {author} {\bibfnamefont {D.~T.}\
  \bibnamefont {Alexander}}, \bibinfo {author} {\bibfnamefont {P.}~\bibnamefont
  {Ghosez}},  \emph {et~al.},\ }\href {\doibase 10.1038/s41563-020-0757-x}
  {\bibfield  {journal} {\bibinfo  {journal} {Nature Materials}\ }\textbf
  {\bibinfo {volume} {19}},\ \bibinfo {pages} {1182} (\bibinfo {year}
  {2020})}\BibitemShut {NoStop}%
\bibitem [{\citenamefont {Zhang}\ \emph {et~al.}(2018)\citenamefont {Zhang},
  \citenamefont {Schmitt}, \citenamefont {Mercy}, \citenamefont {Wang},\ and\
  \citenamefont {Ghosez}}]{ghosez}%
  \BibitemOpen
  \bibfield  {author} {\bibinfo {author} {\bibfnamefont {Y.}~\bibnamefont
  {Zhang}}, \bibinfo {author} {\bibfnamefont {M.~M.}\ \bibnamefont {Schmitt}},
  \bibinfo {author} {\bibfnamefont {A.}~\bibnamefont {Mercy}}, \bibinfo
  {author} {\bibfnamefont {J.}~\bibnamefont {Wang}}, \ and\ \bibinfo {author}
  {\bibfnamefont {P.}~\bibnamefont {Ghosez}},\ }\href {\doibase
  10.1103/PhysRevB.98.081108} {\bibfield  {journal} {\bibinfo  {journal} {Phys.
  Rev. B}\ }\textbf {\bibinfo {volume} {98}},\ \bibinfo {pages} {081108}
  (\bibinfo {year} {2018})}\BibitemShut {NoStop}%
\bibitem [{\citenamefont {Ghosh}\ \emph {et~al.}(2017)\citenamefont {Ghosh},
  \citenamefont {Borisevich},\ and\ \citenamefont {Pantelides}}]{ghosh2017}%
  \BibitemOpen
  \bibfield  {author} {\bibinfo {author} {\bibfnamefont {S.}~\bibnamefont
  {Ghosh}}, \bibinfo {author} {\bibfnamefont {A.~Y.}\ \bibnamefont
  {Borisevich}}, \ and\ \bibinfo {author} {\bibfnamefont {S.~T.}\ \bibnamefont
  {Pantelides}},\ }\href {\doibase 10.1103/PhysRevLett.119.177603} {\bibfield
  {journal} {\bibinfo  {journal} {Phys. Rev. Lett.}\ }\textbf {\bibinfo
  {volume} {119}},\ \bibinfo {pages} {177603} (\bibinfo {year}
  {2017})}\BibitemShut {NoStop}%
\bibitem [{\citenamefont {Rogge}\ \emph
  {et~al.}(2018{\natexlab{b}})\citenamefont {Rogge}, \citenamefont {Green},
  \citenamefont {Shafer}, \citenamefont {Fabbris}, \citenamefont {Barbour},
  \citenamefont {Lefler}, \citenamefont {Arenholz}, \citenamefont {Dean},\ and\
  \citenamefont {May}}]{May}%
  \BibitemOpen
  \bibfield  {author} {\bibinfo {author} {\bibfnamefont {P.~C.}\ \bibnamefont
  {Rogge}}, \bibinfo {author} {\bibfnamefont {R.~J.}\ \bibnamefont {Green}},
  \bibinfo {author} {\bibfnamefont {P.}~\bibnamefont {Shafer}}, \bibinfo
  {author} {\bibfnamefont {G.}~\bibnamefont {Fabbris}}, \bibinfo {author}
  {\bibfnamefont {A.~M.}\ \bibnamefont {Barbour}}, \bibinfo {author}
  {\bibfnamefont {B.~M.}\ \bibnamefont {Lefler}}, \bibinfo {author}
  {\bibfnamefont {E.}~\bibnamefont {Arenholz}}, \bibinfo {author}
  {\bibfnamefont {M.~P.~M.}\ \bibnamefont {Dean}}, \ and\ \bibinfo {author}
  {\bibfnamefont {S.~J.}\ \bibnamefont {May}},\ }\href {\doibase
  10.1103/PhysRevB.98.201115} {\bibfield  {journal} {\bibinfo  {journal} {Phys.
  Rev. B}\ }\textbf {\bibinfo {volume} {98}},\ \bibinfo {pages} {201115}
  (\bibinfo {year} {2018}{\natexlab{b}})}\BibitemShut {NoStop}%
\bibitem [{\citenamefont {Shaikh}\ \emph
  {et~al.}(2020{\natexlab{b}})\citenamefont {Shaikh}, \citenamefont
  {Karmakar},\ and\ \citenamefont {Ghosh}}]{PRB2020}%
  \BibitemOpen
  \bibfield  {author} {\bibinfo {author} {\bibfnamefont {M.}~\bibnamefont
  {Shaikh}}, \bibinfo {author} {\bibfnamefont {M.}~\bibnamefont {Karmakar}}, \
  and\ \bibinfo {author} {\bibfnamefont {S.}~\bibnamefont {Ghosh}},\ }\href
  {\doibase 10.1103/PhysRevB.101.054101} {\bibfield  {journal} {\bibinfo
  {journal} {Physical Review B}\ }\textbf {\bibinfo {volume} {101}},\ \bibinfo
  {pages} {054101} (\bibinfo {year} {2020}{\natexlab{b}})}\BibitemShut
  {NoStop}%
\bibitem [{\citenamefont {Hohenberg}\ and\ \citenamefont {Kohn}(1964)}]{DFT}%
  \BibitemOpen
  \bibfield  {author} {\bibinfo {author} {\bibfnamefont {P.}~\bibnamefont
  {Hohenberg}}\ and\ \bibinfo {author} {\bibfnamefont {W.}~\bibnamefont
  {Kohn}},\ }\href {\doibase 10.1103/PhysRev.136.B864} {\bibfield  {journal}
  {\bibinfo  {journal} {Phys. Rev.}\ }\textbf {\bibinfo {volume} {136}},\
  \bibinfo {pages} {B864} (\bibinfo {year} {1964})}\BibitemShut {NoStop}%
\bibitem [{\citenamefont {King-Smith}\ and\ \citenamefont
  {Vanderbilt}(1993)}]{Berry}%
  \BibitemOpen
  \bibfield  {author} {\bibinfo {author} {\bibfnamefont {R.~D.}\ \bibnamefont
  {King-Smith}}\ and\ \bibinfo {author} {\bibfnamefont {D.}~\bibnamefont
  {Vanderbilt}},\ }\href {\doibase 10.1103/PhysRevB.47.1651} {\bibfield
  {journal} {\bibinfo  {journal} {Phys. Rev. B}\ }\textbf {\bibinfo {volume}
  {47}},\ \bibinfo {pages} {1651} (\bibinfo {year} {1993})}\BibitemShut
  {NoStop}%
\bibitem [{\citenamefont {Kresse}\ and\ \citenamefont
  {Furthm\"uller}(1996)}]{vasp}%
  \BibitemOpen
  \bibfield  {author} {\bibinfo {author} {\bibfnamefont {G.}~\bibnamefont
  {Kresse}}\ and\ \bibinfo {author} {\bibfnamefont {J.}~\bibnamefont
  {Furthm\"uller}},\ }\href {\doibase 10.1103/PhysRevB.54.11169} {\bibfield
  {journal} {\bibinfo  {journal} {Phys. Rev. B}\ }\textbf {\bibinfo {volume}
  {54}},\ \bibinfo {pages} {11169} (\bibinfo {year} {1996})}\BibitemShut
  {NoStop}%
\bibitem [{\citenamefont {Anisimov}\ \emph {et~al.}(1997)\citenamefont
  {Anisimov}, \citenamefont {Aryasetiawan},\ and\ \citenamefont
  {Lichtenstein}}]{DFTU}%
  \BibitemOpen
  \bibfield  {author} {\bibinfo {author} {\bibfnamefont {V.~I.}\ \bibnamefont
  {Anisimov}}, \bibinfo {author} {\bibfnamefont {F.}~\bibnamefont
  {Aryasetiawan}}, \ and\ \bibinfo {author} {\bibfnamefont {A.}~\bibnamefont
  {Lichtenstein}},\ }\href@noop {} {\bibfield  {journal} {\bibinfo  {journal}
  {Journal of Physics: Condensed Matter}\ }\textbf {\bibinfo {volume} {9}},\
  \bibinfo {pages} {767} (\bibinfo {year} {1997})}\BibitemShut {NoStop}%
\bibitem [{\citenamefont {Dudarev}\ \emph {et~al.}(1998)\citenamefont
  {Dudarev}, \citenamefont {Botton}, \citenamefont {Savrasov}, \citenamefont
  {Humphreys},\ and\ \citenamefont {Sutton}}]{LSDAU}%
  \BibitemOpen
  \bibfield  {author} {\bibinfo {author} {\bibfnamefont {S.~L.}\ \bibnamefont
  {Dudarev}}, \bibinfo {author} {\bibfnamefont {G.~A.}\ \bibnamefont {Botton}},
  \bibinfo {author} {\bibfnamefont {S.~Y.}\ \bibnamefont {Savrasov}}, \bibinfo
  {author} {\bibfnamefont {C.~J.}\ \bibnamefont {Humphreys}}, \ and\ \bibinfo
  {author} {\bibfnamefont {A.~P.}\ \bibnamefont {Sutton}},\ }\href {\doibase
  10.1103/PhysRevB.57.1505} {\bibfield  {journal} {\bibinfo  {journal} {Phys.
  Rev. B}\ }\textbf {\bibinfo {volume} {57}},\ \bibinfo {pages} {1505}
  (\bibinfo {year} {1998})}\BibitemShut {NoStop}%
\bibitem [{\citenamefont {Perdew}\ \emph {et~al.}(2008)\citenamefont {Perdew},
  \citenamefont {Ruzsinszky}, \citenamefont {Csonka}, \citenamefont {Vydrov},
  \citenamefont {Scuseria}, \citenamefont {Constantin}, \citenamefont {Zhou},\
  and\ \citenamefont {Burke}}]{PBEsol}%
  \BibitemOpen
  \bibfield  {author} {\bibinfo {author} {\bibfnamefont {J.~P.}\ \bibnamefont
  {Perdew}}, \bibinfo {author} {\bibfnamefont {A.}~\bibnamefont {Ruzsinszky}},
  \bibinfo {author} {\bibfnamefont {G.~I.}\ \bibnamefont {Csonka}}, \bibinfo
  {author} {\bibfnamefont {O.~A.}\ \bibnamefont {Vydrov}}, \bibinfo {author}
  {\bibfnamefont {G.~E.}\ \bibnamefont {Scuseria}}, \bibinfo {author}
  {\bibfnamefont {L.~A.}\ \bibnamefont {Constantin}}, \bibinfo {author}
  {\bibfnamefont {X.}~\bibnamefont {Zhou}}, \ and\ \bibinfo {author}
  {\bibfnamefont {K.}~\bibnamefont {Burke}},\ }\href {\doibase
  10.1103/PhysRevLett.100.136406} {\bibfield  {journal} {\bibinfo  {journal}
  {Phys. Rev. Lett.}\ }\textbf {\bibinfo {volume} {100}},\ \bibinfo {pages}
  {136406} (\bibinfo {year} {2008})}\BibitemShut {NoStop}%
\bibitem [{\citenamefont {Cho}\ \emph {et~al.}(2021)\citenamefont {Cho},
  \citenamefont {Klyukin}, \citenamefont {Ning}, \citenamefont {Li},
  \citenamefont {Comin}, \citenamefont {Green}, \citenamefont {Yildiz},\ and\
  \citenamefont {Ross}}]{cho2021}%
  \BibitemOpen
  \bibfield  {author} {\bibinfo {author} {\bibfnamefont {E.}~\bibnamefont
  {Cho}}, \bibinfo {author} {\bibfnamefont {K.}~\bibnamefont {Klyukin}},
  \bibinfo {author} {\bibfnamefont {S.}~\bibnamefont {Ning}}, \bibinfo {author}
  {\bibfnamefont {J.}~\bibnamefont {Li}}, \bibinfo {author} {\bibfnamefont
  {R.}~\bibnamefont {Comin}}, \bibinfo {author} {\bibfnamefont {R.~J.}\
  \bibnamefont {Green}}, \bibinfo {author} {\bibfnamefont {B.}~\bibnamefont
  {Yildiz}}, \ and\ \bibinfo {author} {\bibfnamefont {C.~A.}\ \bibnamefont
  {Ross}},\ }\href@noop {} {\bibfield  {journal} {\bibinfo  {journal} {Physical
  Review Materials}\ }\textbf {\bibinfo {volume} {5}},\ \bibinfo {pages}
  {094413} (\bibinfo {year} {2021})}\BibitemShut {NoStop}%
\bibitem [{\citenamefont {Sivadas}\ \emph {et~al.}(2021)\citenamefont
  {Sivadas}, \citenamefont {Doak},\ and\ \citenamefont {Ganesh}}]{sivadas2021}%
  \BibitemOpen
  \bibfield  {author} {\bibinfo {author} {\bibfnamefont {N.}~\bibnamefont
  {Sivadas}}, \bibinfo {author} {\bibfnamefont {P.}~\bibnamefont {Doak}}, \
  and\ \bibinfo {author} {\bibfnamefont {P.}~\bibnamefont {Ganesh}},\ }\href
  {https://arxiv.org/abs/2106.08783} {\bibfield  {journal} {\bibinfo  {journal}
  {arXiv preprint arXiv:2106.08783}\ } (\bibinfo {year} {2021})}\BibitemShut
  {NoStop}%
\end{thebibliography}
%
\end{document}